\documentclass[manuscript]{acmart}
\usepackage{dirtytalk}
\usepackage{color}

\newcommand{\new}[1]{\textcolor{black}{#1}}

\AtBeginDocument{%
  \providecommand\BibTeX{{%
    \normalfont B\kern-0.5em{\scshape i\kern-0.25em b}\kern-0.8em\TeX}}}

\begin{document}

\title{Should Young Computer Scientists Stop Collaborating with their Doctoral Advisors?}

\author{Ariel Rosenfeld}
\email{ariel.rosenfeld@biu.ac.il}
\orcid{0000-0002-3230-3060}
\author{Oleg Maksimov}
\email{oleg@maksimov.co.il}
\affiliation{%
  \institution{Bar-Ilan University}
  \streetaddress{Max and Anna Web}
  \city{Ramat Gan}
  \country{Israel}
  \postcode{5290002 }
}

\renewcommand{\shortauthors}{Rosenfeld and Maksimov}

\begin{abstract}
 {One of the first steps in an academic career, and perhaps the pillar thereof, is completing a PhD under the supervision of a doctoral advisor. While prior work has examined the advisor-advisee relationship and its potential effects on the prospective academic success of the advisee, very little is known on the possibly continued relationship after the advisee has graduated. We harnessed three genealogical and scientometric datasets to identify 3 distinct groups of computer scientists: Highly independent, who cease collaborating with their advisors (almost) instantly upon graduation; Moderately independent, who (quickly) reduce the collaboration rate over $\sim$5 years; and Weakly independent who continue collaborating with their advisors for at least 10 years post-graduation. We find that highly independent researchers are more academically successful than their peers in terms of H-index, i10-index and total number of citations throughout their careers. Moderately independent researchers perform, on average, better than weakly independent researchers, yet the differences are not found to be statistically significant. In addition, both highly and moderately independent researchers are found to have longer academic careers. \new{Interestingly,} weakly independent researchers tend to be  supervised by more academically successful advisors.}
\end{abstract}



\maketitle

\section{Introduction}

Shortly after the first author started his tenure-track position at Bar-Ilan University, he published a few additional papers with his doctoral advisor. These papers were mostly \say{lingering} results from his PhD or direct extensions thereof. He was very surprised that his Department Chair reprimanded him for this, claiming that it could be harmful to his career. Surprisingly, until now, we were unable to find any support to that claim in the literature. 

The benefits and importance of mentoring have been long established and span a wide variety of vocational fields both in and outside of academia \cite{kram1988mentoring,eby2008does}.   
In the academic realm, the supervision benefits are commonly mutual \cite{johnson2015being}: the advisor extends her ability to conduct research by delegation, extends her influence network, etc. and the advisee learns the important skills needed to conduct scientific research, receives various types of academic support, etc. 
Focusing on the advisee, prior research has shown that the doctoral advisor's identity and characteristics can have a far reaching effect on a doctoral student's future career. For example, having an advisor with a strong publication record was shown to drive graduate students' publication activity \cite{qi2017standing}, to increase students' chances of obtaining an academic position \cite{lienard2018intellectual} and to serve as a predictor for future academic success \cite{li2019early}. These, in turn, include higher levels of scientific autonomy \cite{horta2016impact}, active international collaboration dynamics \cite{amara2015can}, an increase in the advisee's chances of pioneering their own research topics (i.e., not following their advisor's research topics), winning prestigious prizes and recognition \cite{ma2020mentorship} and publishing in top venues such as Nature and Science \cite{sekara2018chaperone}. 
An advisor's strong publication record is not the only factor to influence the advisee's career trajectory. A satisfactory advisor-advisee relationship is an essential component of successful doctoral training \cite{zhao2007more,styles2001synergistic,wisker2007postgraduate,taylor2005handbook,lee2019successful},  the number of advisors and their characteristics \cite{wuestman2020genealogical}, the number of advisees that an advisor mentors \cite{malmgren2010role} and the advisor's academic age \cite{liu2018understanding} all influence the advisee's future academic path, to name a few. However, to the best of our knowledge, existing literature has yet to focus on the possibly \textit{continued relationship after the advisee has completed her doctoral studies}. One notable exception is  \cite{ma2020mentorship}, who investigated various possible effects of having an award-winning advisor on the future success of the advisee across various fields (but not in computer science). The authors found that the proportion of co-authored papers between an advisee and her advisor within the advisee's total body of work negatively relates to her chances of becoming an award winner herself.  This result may suggest that a continued advisor-advisee relationship indeed bears significant (negative) effects on the advisee's academic success in her career.

\new{In this work, we adopt a more fine-grained approach than \cite{ma2020mentorship} which allows us to reach more nuanced conclusions for computer science. Specifically, we examine richer measures of academic success (also known as impact metrics) rather than awards, and follow the advisor-advisee collaboration pattern on a yearly basis rather than relying on overall frequency of co-authored papers in one's body of work. Overall, our results confirm and significantly extend those reached in \cite{ma2020mentorship} by identifying  three distinct advisor-advisee collaboration patterns in computer science, which are in turn linked to the advisees' academic success and the advsiors' characteristics. }

\section{Methods}



The data used in this study comes from three highly popular datasets:  DBLP\footnote{\url{https://dblp.uni-trier.de/}} ($\sim5.5M$ papers and $\sim1.7M$ authors), Academic Family Tree (AFT)\footnote{\url{https://academictree.org/}} ($\sim700K$ authors) and Microsoft Academic Graph (MAG) ($\sim36M$ papers and $\sim18M$ authors)\footnote{\url{https://academic.microsoft.com/}, \new{no longer active since January 2022}.}.

DBLP is the most popular computer science database today which indexes all major CS journals, conference proceedings, books and preprint servers.
From DBLP, we extracted all indexed data up to 2020 and focused on all authors who had published at least 5 papers over a publication career span of at least 5 years (as was done in past works e.g., \cite{liu2018understanding}), resulting in over half a million authors who published approximately $12.5M$ \textit{non-unique} papers (average of 22 papers per author) and over half a million \textit{unique} papers. 

AFT is a crowdsourced academic genealogy database which is often used in academic research to investigate different aspects of advisor-advisee relationships, e.g.,  \cite{sanyal2020gm,lienard2018intellectual,ma2020mentorship}. \new{While AFT may not be completely representative of the advisor-advisee population, it does provide comprehensive state-of-the-art indexing of a variety} of academic supervision relationships, including PhD supervisions, along with basic (mostly partial) information on the first and \new{last} years of the supervision. From AFT, we extracted all advisor-advisee pairs for which a PhD supervision was indicated and filtered out any pair for which the advisee or advisor was absent from the extracted DBLP author pool, if both the start and last supervision years were missing or if the advisee had less than five active publication years after her graduation.
We completed the missing first or last years as suggested in \cite{lienard2018intellectual}. Specifically, we inferred the missing year by identifying the earliest commonly authored publication by the advisor and advisee and added or subtracted the median lag between the start/end year and first publication (in our data, $4$ years is the median PhD duration and $1$ year is the median duration from start to first publication). This process resulted in $\sim14K$  advisor-advisee pairs for which complete estimated data is derived.\footnote{We have examined the validity of using such data completion techniques in our data and noticed that it brings about extremely similar results to those derived solely based on the $1K$ pairs for which all information is given.} Out of these $\sim14K$  pairs, $3,401$ advisees have at least 5 active publication years \textit{after} their PhD graduation, and $993$ have at least 10 active years after graduation. Since only $236$ advisees have at least 15 active years after their PhD graduation, we only consider the entire set of advisees and the former two subsets in our analysis. 

Microsoft Academic Graph (MAG) was used to extract the citations for the papers in our analysis. MAG was recently established as a leading citation database \cite{martin2021google}, which is only second to Google Scholar. Unfortunately, Google Scholar provides very limited access to its data and thus could not be used for our relatively large scale analysis. The majority of papers in our study were matched in MAG and the rest (mainly workshop papers and preprints) were omitted from further consideration.



\section{Results}

\begin{figure}[ht]
\centering
\includegraphics[width=0.4\textwidth]{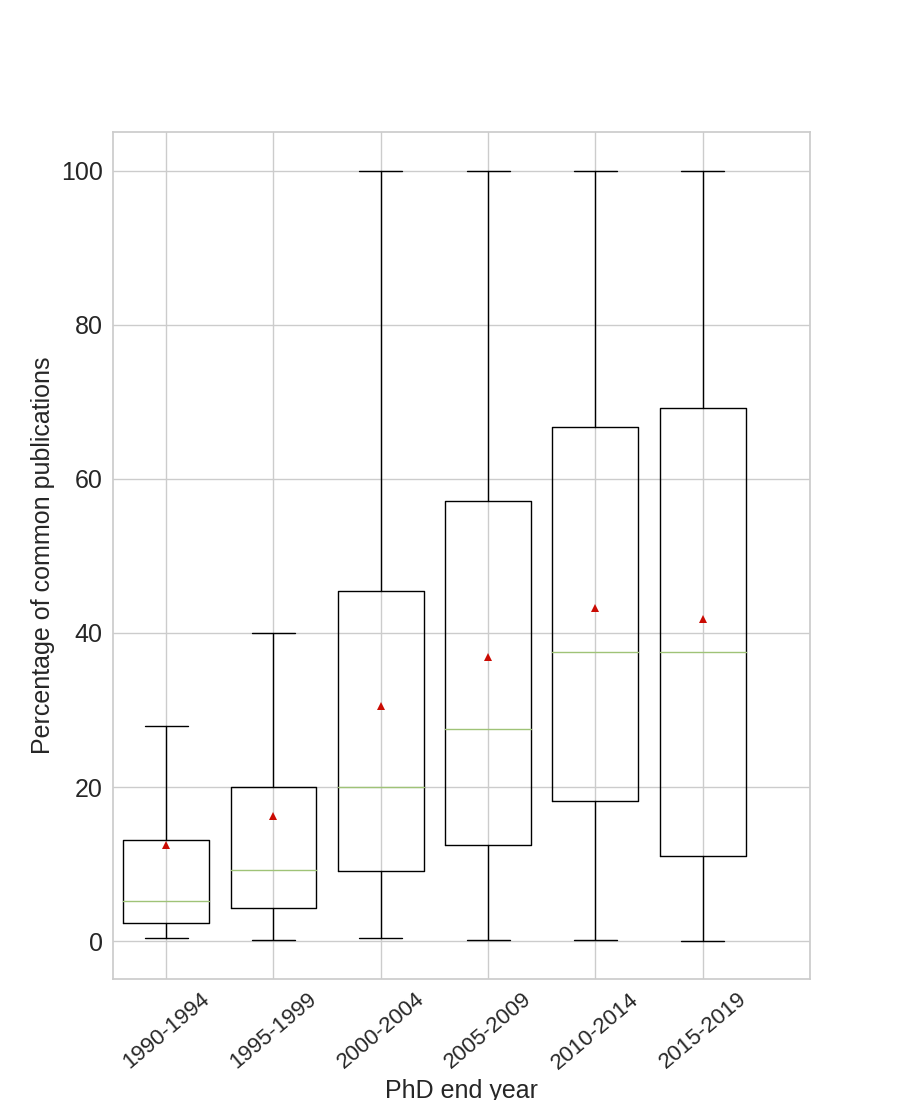}
\caption{PhD graduation year (X-Axis) vs the ratio of co-authored publications with one's advisor (Y-Axis).}
\label{fig:year-pubs}
\end{figure}

We start by examining the nature of the continued advisor-advisee relationship by looking at the  \new{portion of advisor-advisee co-authored  papers within the advisee's total body of work} over time as seen in Figure \ref{fig:year-pubs}. 

As can be seen in Figure \ref{fig:year-pubs}, \new{the percentage of advisor-advisee co-authored papers within an advisee's total body of work is growing quickly over time. Namely, more recent graduates tend to publish more with their PhD advisors, in relative terms, compared to prior graduates. In addition, we encounter higher variability in the portion of co-authored papers within one's body of work for more recent graduates.}

\begin{figure}[ht]
\centering
\includegraphics[width=0.33\textwidth]{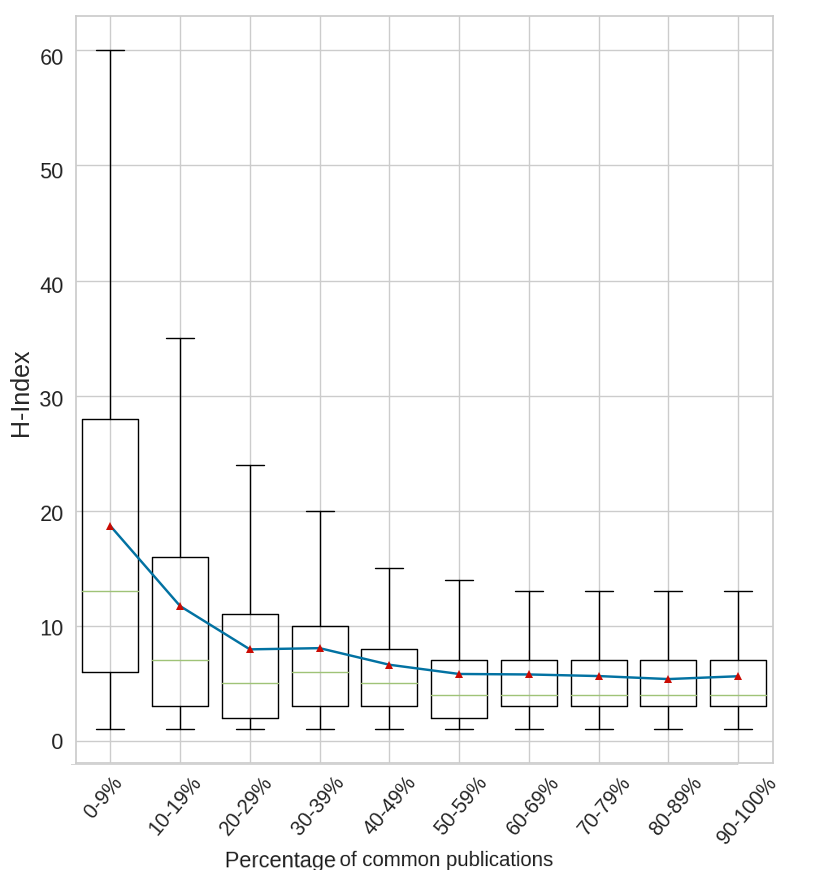}
\includegraphics[width=0.33\textwidth]{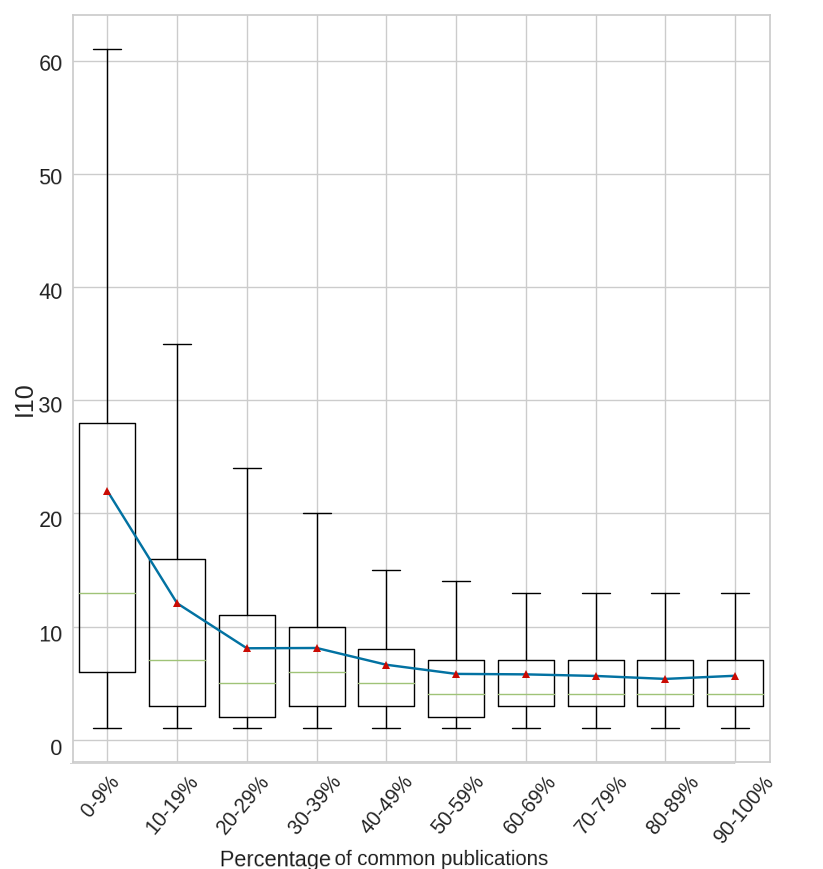}
\includegraphics[width=0.33\textwidth]{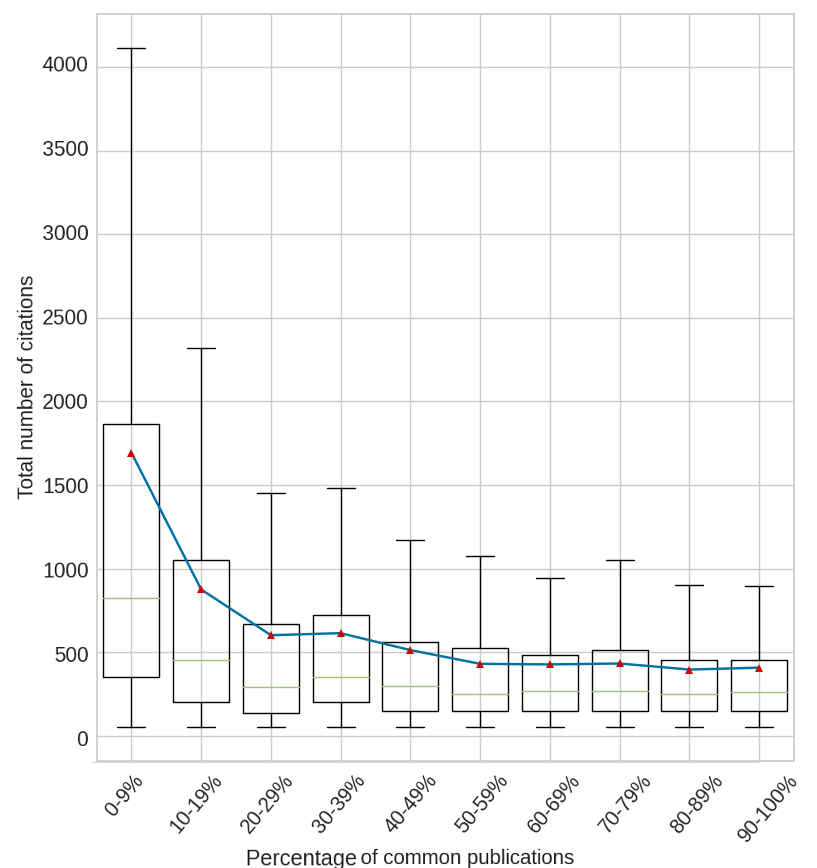}
\caption{Rate of advisor-advisee co-authored papers in an advisee's career (X-Axis) vs H-index (Left), i10-index (Middle) and total number of citations (Right).}
\label{fig:pubs-metrics}
\end{figure}

\new{Next,} we extend and verify the results from \cite{ma2020mentorship} (given for other fields of science and different success metrics) and examine the portion of advisor-advisee  co-authored publications within the advisee's total body of work against three classic measures of academic success 5 and 10 years after graduation. These measures are: H-index, i10-index and total number of citations (see \cite{waltman2016review} for a review of these academic metrics and their use in practice). 
\new{As can be seen in Figure \ref{fig:pubs-metrics},  higher collaboration rates are linked with lower academic success metrics. The three success metrics seem to present a very similar  exponentially decaying behavior along the collaboration frequency axis. For concreteness, the mean H-Index, i10 and total number citations drop by 61\%-71\% as the collaboration frequency is increased from the 0-9\% range to 40-49\%. Similarly, the variability in each success metric is quickly reduced as the collaboration frequency increases. }   


Combining the results above, we see that, over time, younger computer scientists seem to publish more frequently with their advisors after their graduation, which in turn is linked with lower academic success. 
However, the above analysis misses what we consider to be important pieces of the puzzle - what collaboration \textit{patterns} exist and how these are connected with academic success. 

\begin{figure}[ht]
\centering
\includegraphics[width=0.55\textwidth]{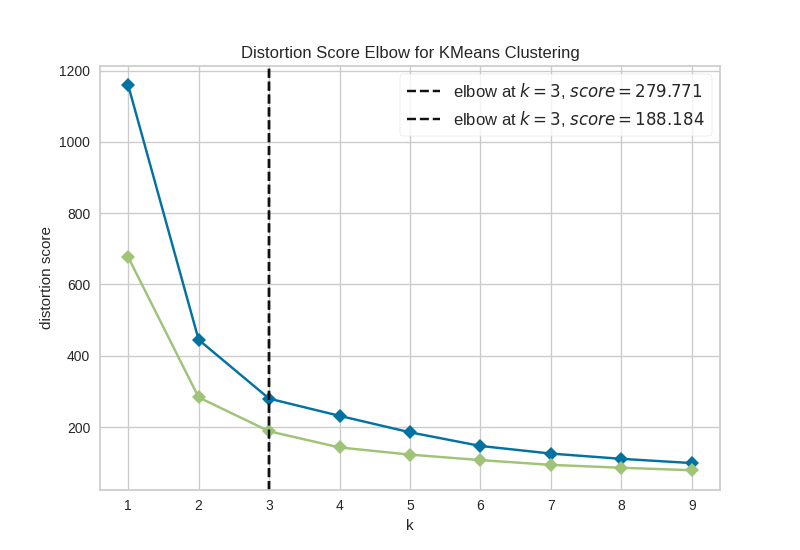}
\caption{Elbow graph for determining optimal $k$ in clustering the collaboration patterns. The X-Axis represents the number of clusters ($k$) and the Y-Axis represents the distortion (variance) in the data. The top line (blue) is for the 5-year time span after graduation, followed by the line (green) for the 10-year time span.}
\label{fig:elbow}
\end{figure}

To that end, we employ a cluster analysis while focusing on researchers with sufficiently long careers after their graduation (i.e., researchers with at least 5 and 10 active publication years after their PhD graduation).
Specifically, we look at the ratio of papers each researcher has co-authored with her PhD advisor \textit{per year} for the first $5$ years after graduation and the first $10$ years after graduation and seek to group these time series into meaningful groups (or clusters) which could be analysed and compared. We adopt the popular K-means algorithm \cite{macqueen1967some} with Dynamic Time Warping (DTW) distance measure \cite{esling2012time}, which is especially suited for time series clustering. 
In order to determine the appropriate number of clusters ($k$), we use the classic Elbow Method (which can be traced back to the mid-20th-century~\cite{thorndike1953belongs}).
%
As can be seen from Figure \ref{fig:elbow}, a reasonable selection of $k$, under both time spans we examined, is $k=3$. 

\begin{figure}[H]
\centering
\includegraphics[width=0.48\textwidth]{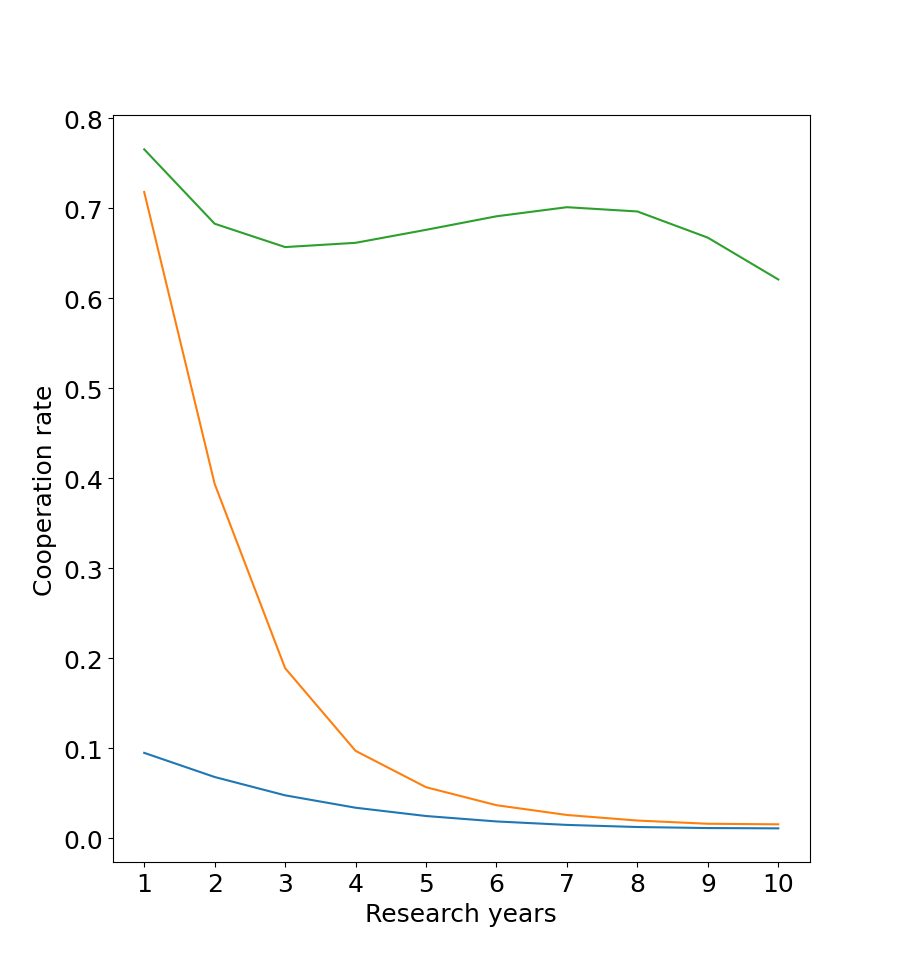}
\caption{Cluster centroids as representatives of the three groups of researchers \new{10 years after graduation}. The X-Axis represents time after graduation and the Y-Axis represents the portion of co-authored papers the researcher has with her advisor each year. \new{\textit{Highly Independent} researchers are represented by the bottom line in blue, \textit{Moderately Independent} researchers are represented by the middle line in orange, and \textit{Weakly Independent} researchers are represented by the top line in green.}}
\label{fig:clusters}
\end{figure}

The three clusters identified for the 10-year time span after graduation are depicted in Figure \ref{fig:clusters} using the centroid of each cluster \new{(the diagram for the 5-year time span after graduation is almost identical to the first 5 years of each cluster presented in the figure)}. As can clearly be seen, three distinct patterns arise: 1) \textit{Highly Independent} researchers who (almost) instantly stop collaborating with their advisors upon graduation; 2)  \textit{Moderately Independent} researchers who gradually stop collaborating with their advisors (over $\sim$5 years); and 3) \textit{Weakly Independent}  researchers who maintain a high degree of collaboration with their advisors. 

For concreteness, let us consider three well-known computer scientists.
\textit{Andrew Ng}, a pioneer in Machine Learning and AI, is one of the leading computer scientists today both in academia and industry. He is a prime example of the highly independent cluster, showing zero co-authored publications with his PhD advisor (Michael I. Jordan) after his graduation in 2013. 
At the other extreme, let us consider \textit{Zhu Han} from University of Houston. He has been in the top 1\% of the most cited researchers in all fields of science according to Web of Science\footnote{\url{https://www.webofscience.com/}} since 2017. Since his graduation in 2003, he has published extensively with his PhD advisor (K.J. Ray), publishing almost 50 papers together. Only in 2007 (4 years after his graduation) did he author his first paper without his advisor, while more than 50\% of his publications that year were with him. 
As a representative of the moderately independent cluster, we consider another advisee of K.J. Ray's, \textit{Wade Trappe}, a world leader in cybersecurity and communication systems from Rutgers University. Since his graduation in 2002, he has gradually ceased collaborating with his advisor -- in 2003 his advisor co-authored 7 out of his 8 papers, followed by 4 out of 8 papers with his advisor in 2004 and 2 out of 9 papers in 2005. Since 2006 he has not published anything with his advisor.

We compare the three clusters in terms of their mean success metrics using a one-way ANOVA (ANalysis Of VAriance) with post-hoc Tukey HSD (Honestly Significant Difference) correction, see Table \ref{tab:h-index}. 
We see that for the three examined success metrics, the highly independent cluster scores statistically significantly higher than the other two clusters both 5 and 10 years after graduation (one exception is the i10-index at 10 years where the clusters do not differ significantly, \new{partially due to the low number of members in the weakly independent cluster}). Surprisingly, while the moderately independent cluster indeed scores higher on all metrics and across both time points compared to the weakly independent cluster, the differences are not found to be statistically significant at $p<0.05$ ($p$ value ranged from 0.2 to 0.5). While the lack of significance for the 10 year time span may be partially attributed to the relatively low number of members in the weakly independent cluster, we cannot provide such an explanation for the 5-year time span. 

Furthermore, most researchers in our data are members of the highly independent cluster (57.3\% and 60.3\% for both the 5 and 10 years after graduation, respectively) or the moderately independent one (34\% and 36.7\%, respectively). The fewest number of researchers are members of the weakly independent cluster (8.7\% and 3\%). This result seems to suggest that more independent graduates are also more likely to \say{survive} in academic research, especially the transition from 5 years to 10 years after graduation. Incidentally or not, this is usually the time frame in which tenure-track researchers in academia apply for a tenured post. To verify our observation we compare the average career length of researchers in each cluster (of the 5-year clustering) using yet another one-way ANOVA with post-hoc Tukey correction. As expected, we find that the clusters significantly differ, with the weakly independent cluster having a significantly \textit{shorter} average career length than the other two clusters at $p<0.05$, demonstrating about 10\% shorter careers. The highly independent and moderately independent clusters do not differ significantly in this respect.

	\begin{table} [H]\centering
	\begin{tabular}{c | l |l |l}
		Mean H-index|i10-index|total citations (N) &  Highly independent &  Moderately independent & Weakly independent\\ [0.5ex]
		\hline
	\textsc{5 years}& \textbf{24}|\textbf{28}|\textbf{2311}(1950) &22|25|1924 (1155) & 20|22|1622 (296)   \\
	\hline	
	\textsc{10 years}&  \textbf{38}|{50}|\textbf{4520}  (599) & 36|{47}|3682 (364) & 32|33|2541 (30)\\
		\hline
	\hline
	\end{tabular}
		\caption{Clusters' \new{rounded} mean H-index|i10-index|total citations (size of the cluster in parentheses) at different points in their careers measured by years after their graduation. Results in bold are statistically significantly higher at $p<0.05$ using a one-way ANOVA  with post-hoc Tukey correction.}
	\label{tab:h-index}
	\end{table}

The identified clusters seem to differ in their members' collaboration behavior with their advisors even in the very first years after graduation and significantly differ in their academic success and their chances of having long academic careers. However, one may also wonder if the clusters also differ \textit{before \new{and at}} graduation. We examine the following key metrics: the number of papers that the advisee published during her PhD period (with and without the advisor), the advisor's academic age (i.e., time since first publication), the advisor's performance metrics (H-index, i10-index and total citations), \new{the advisor's country of affiliation (based on Scopus\footnote{\url{https://www.scopus.com/}})\ and the advisee's graduation year}. 

We find that the highly and weakly independent cluster members publish significantly \textit{more} than the moderately independent cluster members during their PhD, $p<0.05$. Specifically, the moderately independent researchers publish 3.45 papers, on average, during their PhD period while the highly and weakly independent researchers average 4.08 and 3.95, respectively. In other words, highly and weakly independent researchers do not differ significantly in their academic productivity during their PhD period, despite their considerable differences after graduation, yet they are both more productive than moderately independent researchers, on average.   
Be that as it may, highly independent researchers do demonstrate significantly higher levels of academic independence also during their PhD period. On average, a highly independent researcher publishes 49\% of her papers with her advisor during her PhD period compared to 61\% and 69\% in the moderately and weakly independent clusters, respectively. The latter two groups do not differ significantly. In other words, a researcher's independence post-graduation is strongly linked with her independence during her PhD.
Focusing on the advisors' academic status and success, we find that the highly and moderately independent researchers were advised by advisors of \say{similar caliber} (i.e., no significant difference in the examined success metrics) and, as such, are expected to attract students of similar academic potential and promise. \new{Interestingly}, we find that \textit{weakly independent researchers} are supervised by significantly \textit{more academically successful advisors} in terms of both H-index and i10-index, $p<0.05$. On average, weakly independent researchers were advised by advisors with 9-10\% higher H-indexes and 14-16\% higher i10-index compared to the other two groups. No significant difference was found for the total number of citations metric. \new{We further find that weakly independent researchers graduate slightly later than the members of the other two groups by an average of 1.5 years, $p<0.05$. Highly and moderately independent researchers do not differ on this account.}
No statistically significant differences were found for the advisors' academic age \new{ and the distribution over country of affiliation for all three groups.}



\section{Discussion}

Our analysis of genealogical and scientometric databases reveals three distinct collaboration patterns between computer science doctoral graduates and their advisors:  \textit{Highly Independent} who cease to collaborate with their advisors almost instantly upon graduation; \textit{Moderately Independent} who gradually stop the collaboration over $\sim$5 years; and \textit{Weakly Independent} who continue to heavily collaborate with advisors, at least for the first 10 years after their graduation. 
In turn, highly independent researchers are positively linked with greater academic success in terms of H-index, i10-index and total number of citations both 5 and 10 years after graduation. Moderately independent researchers, who also stop collaborating with their doctoral advisors similarly to the first group but at a slower pace, are not found to be more academically successful than the weakly independent researchers, albeit displaying higher average metrics on all accounts. Both highly and moderately independent researchers are also associated with greater chances of having a long academic career. 

Distinguishing between the identified groups seems to be possible even by observing the very first years (1-2) after the advisee graduated. In the same manner, it turns out that the identified groups also differ in their PhD profiles. Most notably, as one could expect, highly independent graduates are also associated with higher levels of independence during their PhD period. \new{Interestingly}, weakly independent researchers tend to be supervised by \textit{more} academically successful advisors. \new{One may} consider this result to be \new{somewhat} surprising, since more academically successful advisors are expected to attract promising doctoral students \new{who are more capable of conducting independent research. However, one may also argue that advisors who continue to publish papers with their former students are effectively expanding their workforce and thus become more successful. Either way,} we speculate that advisees of highly successful advisors may hold the common belief that they may be the next rising stars of their advisor's hit research topic \cite{malmgren2010role}, and thus continue to collaborate with their advisors. Additionally, it may be that highly successful advisors have greater academic resources (e.g., funding, equipment, ideas, collaboration network) that young graduates feel can still assist them after graduation in building their careers. A complete inquiry into this matter is outside the scope of this work and is left for future work. 

Arguably, we interpret the fact that highly and moderately independent researchers are statistically indistinguishable in terms of their advisors' examined characteristics (i.e., age, success metrics \new{and country of affiliation) and in their graduation year} to be supporting evidence of these researchers' similar potential and promise. If this assumption holds, the differences between the groups should be, at least partially, attributed to the collaboration between one and her advisor during and after her doctoral studies. 


Our results provide supporting evidence to the conclusion that young computer scientists should be reasonably encouraged to stop collaborating with their doctoral advisors. The sooner - the better. This may be especially important for productive PhD graduates and those who were supervised by highly successful PhD advisors. It further seems that promoting student independence \textit{during the PhD period} should be encouraged by the PhD advisors. 
Our conclusion is further strengthened by recent evidence outside the computer science realm demonstrating how graduates who undergo additional training outside their advisors' research focus \cite{lienard2018intellectual} and those who change their research agenda away from their advisor's focus \cite{YU2021101193,ma2020mentorship} tend to significantly increase their scientific impact. Clearly, the above translate into lower collaboration rates with one's advisor early on in one's career.

\new{As is the case in most literature on the academic labor markets, little is known about graduates who had very short academic career spans (in our study, less than 5 years after graduation). Specifically, we cannot easily determine why a graduate has ceased to publish academic research (e.g., Has she ever looked for a research post? Did she find one but was later laid off or quit?). As one could expect, we do find that those who qualified for our analysis had published 18\% more papers during their PhD compared to those who did not. We intend to investigate this population in future work. In addition,  }
we plan to examine more subtle aspects of the advisor-advisee relationship including \say{informal} collaboration and advice which need not necessarily result in co-authored papers. Note that this work explicitly assumes that collaboration is manifested in co-authored papers, yet in the academic context and especially between doctoral advisors and their advisees, collaboration and advisement are much more than simple co-authorship \cite{garmire2021mentorship}. Last, we plan to investigate the collaboration patterns between other \say{family members} in the academic genealogy. For example, we plan to extend our analysis and investigate the possible relationships between \say{academic siblings}, i.e., researchers who were advised by the same doctoral advisor. 

\bibliographystyle{ACM-Reference-Format}
\bibliography{MyBib}
\end{document}